\documentclass{article}
\usepackage{fleqn, espcrc2}

\def\fun#1#2{\lower3.6pt\vbox{\baselineskip0pt\lineskip.9pt
\ialign{$\mathsurround=0pt#1\hfil##\hfil$\crcr#2\crcr\sim\crcr}}}
\newcommand{\be}{\begin{equation}}
\newcommand{\ee}{\end{equation}}
\newcommand{\ba}{\begin{eqnarray}}
\newcommand{\ea}{\end{eqnarray}}
\newcommand{\bc}{\begin{center}}
\newcommand{\ec}{\end{center}}
\begin{document}

\title{\noindent Pentaquarks in string dynamics}
\author{I. M. Narodetskii$^{(a)}$, C. Semay$^{(b)}$, B. Silvestre-Brac$^{(c)}$, and
Yu.A.Simonov$^{(a)}$\\
$^{(a)}$Institute of Theoretical and Experimental Physics, 117218 Moscow, Russia \\
$^{(b)}$Groupe de Physique Nucl\'{e}aire Th\'{e}orique,
Universit\'{e} de Mons-Hainaut, Acad\'{e}mie Universitaire
Wallonie-Bruxelles, Place du Parc 20, B-7000 Mons, Belgium \\
$^{c}$Laboratoire de Physique Subatomique et de Cosmologie, Avenue
des Martyrs 53, F-38026 Grenoble-Cedex, France}

\begin{abstract}
The masses of $uudd\bar s $, $uudd\bar d$, and $uuss\bar d$
pentaquarks are evaluated in the framework of  both the Effective
Hamiltonian approach to QCD and spinless Salpeter equation using
the Jaffe-Wilczek diquark approximation and the string interaction
for the diquark-diquark-antiquark system. The masses of the light
pentaquarks are found  to be in the region above 2 GeV. Similar
calculations yield the mass of $[ud]^2\bar c$ pentaquark $\sim$
3250 MeV and $[ud]^2\bar b$ pentaquark $\sim$ 6509 MeV.

\end{abstract}
\maketitle
\section{Introduction}

Recently LEPS \cite{Nakano:2003} and DIANA \cite{Barmin:2003}
Collaborations reported the observation of a very narrow peak in the
$K^+n$ and $K^0p$ invariant mass distribution (called $\Theta^+$),
the existence of which has been confirmed by several experimental groups
in various reaction channels.
These experimental results were motivated by a pioneering paper on
chiral soliton model ($\chi$SM) \cite {Diakonov:1997}. The
reported mass determinations for the $\Theta^+$ are very
consistent, falling in the range 1540$\pm$10 MeV, with a width
smaller than the experimental resolution of $20$ MeV for the
photon and neutrino induced reactions, and of 9 MeV for the ITEP
K$^+$Xe$\to$K$^0$pXe$'$ experiment.

In October 2003, the NA49 Collaboration  at CERN SPS \cite{NA49}
announced evidence for an additional  narrow $\Xi^-\pi^-$
resonance called $\Xi^{--}$(1860) with $I=3/2$,  a mass
 $1862\pm 2$ MeV, and a width below the detector resolution
of about 18 MeV. NA49 also reports evidence for a resonance
$\Xi^0$(1860) decaying into $\Xi^-(1320)\pi^+$.  More recently,
H1 collaboration at HERA \cite{Aktas:2004qf} observed a narrow
resonance $ \Theta_c(3099) $ in $D^{* \, -} p$ and $D^{* \, +}
\bar{p}$ invariant mass combinations with a mass $ 3099\pm 6 $
MeV and a decay width $ 12\pm 3 $ MeV.

The complete list of both positive and negative experimental
results on $\Theta^+(1540)$, $\Xi^0$(1860), and  $ \Theta_c(3099)
$ existing by May 2004 is given, {\it e.g.}, in \cite{HSQCD}.
While a dozen experiments have reported evidence for the
phenomenon, another dozen haven't seen the states. In particular,
the $e^+e^-$ collider data (Belle, BaBar, ALEPH, DELPHI) give null
results. The Fermilab high-energy proton experiments do not
observe $\Xi(1860)$, $\Theta^+$, and $\Theta_c$ (CDF), do not
confirm $\Theta_c$ in both $D^{*-}p$ and $D^-p$ channels (FOCUS),
and do not observe $\Theta\to pK_S^0$ (HyperCP, E690)
\cite{Litvintsev}.

Therefore, the existence of the pentaquark is, at the present
time, an {\it experimental} issue.   New results are expected in
the near future: CLAS high statistics proton data (just done),
HERMES with the double statistics (now running), COSY-TOF (five
times more statistics by the end of 2004), KEK  $\pi^+p\to
K\Theta^+$ experiment E359 (will run in May 2005), and some
others.

From the soliton point of view, the $\Theta^+$ is not exotic as
compared to other baryons - it is just a member of
${\bf\overline{10}}_F$  multiplet with $S=+1$. However, in the
sense of the quark model, the $\Theta^+(1540)$ baryon with
positive amount of strangeness is manifestly exotic -- its minimal
configuration can not be satisfied by three quarks. The positive
strangeness requires a $\bar s$ antiquark and $qqqq$ (where $q$
refers to the lightest quarks $u,d$) are required for the net
baryon number, thus making a pentaquark $uudd\bar s$ state as the
minimal ``valence'' configuration. Similarly, $\Theta_c$ can be
interpreted as a pentaquark baryon with minimal quark content $
(udud \bar c) $; thus it is the first exotic baryon with an
anti-charm quark, implying the existence of other exotic baryons
with heavy quarks. The experimental evidence for the manifestly
exotic baryon states provides an opportunity to refine our
quantitative understanding of nonperturbative QCD at low energy.

In this talk, we explore the phenomenology of, and models for, the
exotic baryons, the lowest of which is $\Theta^+(1540)$. The next
section includes a brief review of the theoretical expectations
for the masses of exotic baryons. In section 3, we provide an
overview of the Effective Hamiltonian approach to QCD
\cite{lisbon} and, in section 4, pay  special attention to the
dynamical calculation of pentaquark masses in the framework of
this approach. Section 5 contains some concluding remarks.

\section{Review of Theoretical Estimations}
The $\Theta^+$-hyperon has hypercharge $Y=2$ and third component
of isospin $I_3=0$. The apparent absence of $I_{3}=+1$,
$\Theta^{++}$ in the $K^{+}p$ channel argues against $I=1$;
therefore it is usually assumed the $\Theta^+$ to be an
isosinglet.
The other quantum numbers are not established yet.

As to the theoretical predictions, we are faced with a somewhat
ambiguous situation, in which exotic baryons {\it may} have been
discovered, but there are important controversies with theoretical
estimations for masses of pentaquark states. The experimental
results triggered vigorous theoretical activity and put renewed
urgency in the need to understand of how baryon  properties are
obtained from QCD.

As of today, the most successful prediction comes from the $\chi$SM
\cite{Diakonov:1997}. Recently Ellis {\it et al.} ~\cite{EKP}
critically discussed the calculation of the $\Theta^+$ and
$\Xi^{--}$ masses (and widths) in the $\chi$SM and explored the
unbiased theoretical and phenomenological uncertainties. Overall,
they found the ranges
$1432~{\rm MeV} < m_{\Theta^+} < 1657~{\rm MeV}$,
$1786~{\rm MeV} < m_{\Xi^{--}} <
1970~{\rm MeV}$.
These ranges certainly include the observed masses
$m_{\Theta^+}{=}1539{\pm}2$~{\rm MeV} and $m_{\Xi^{--}} = 1862 \pm
2$~{\rm MeV}, but more precise predictions cannot be made without
introducing more assumptions.

All other attempts of theoretical estimations of the pentaquark
masses can be subdivided into the following five categories: i)
phenomenological analyses of the SU(3)$_F$ mass relations, ii)
phenomenological analysis of the hyperfine splitting in the
Sakharov-Zeldovich (SZ) quark model in which the hadron mass is
just the sum of the effective masses $m_i$ of its constituents
plus hyperfine interaction; iii) dynamical calculations using the
sum rules or lattice QCD, iv) quark model (QM) calculations, and
v) dynamical calculations using the chiral SU(3) quark model.

The QCD sum rules predict a negative parity $\Theta^+$ of mass
$\simeq 1.5$ GeV, while no positive parity state was
found~\cite{sumrules}. The lattice QCD calculations have claimed a
pentaquark signal of either negative parity \cite{Fodor},
\cite{Sasaki}, or positive parity \cite{lattice_china}, in the
vicinity of 1540 MeV. However, the most recent quenched
calculation \cite{lattice_mathur} seems to reveal no evidence for
a pentaquark state with quantum numbers
$I(J^P)=0(\frac{1}{2}^{\pm})$ near this mass region.

The naive quark models, in which all constituents are in a
relative S-wave, naturally predict the ground state energy of a
$J^P=\frac{1}{2}^-$ pentaquark to be lower than that of a
$J^P=\frac{1}{2}^+$ one. Using the arguments based on both the
Goldstone boson exchange between constituent quarks and
color-magnetic exchange it was mentioned \cite{stancu} that the
increase of hyperfine energy in going from negative to positive
parity states can be quite enough to compensate the orbital
excitation energy $\sim 200-250$ MeV. Existing dynamical
calculations of pentaquark masses using the chiral SU(3) quark
model (see {\it e.g.} \cite{chiral_quark_model}) predict the mass
of the $J^P=\frac{1}{2}^+$ pentaquark in the region 2 GeV.
However, the model itself is subject to significant uncertainties
and the results can not be considered as conclusive.

Pentaquark baryons are unexpectedly light. Indeed, a naive QM with
quark mass $\sim$ 350 MeV predicts $\Theta^+$ at about
350$\times$5=1750 MeV plus $\sim$150 MeV for strangeness
plus$\sim$300 MeV for the P-wave excitation. A natural remedy
would be to decrease the number of constituents. This leads one to
consider dynamical clustering into subsystems of diquarks like
$[ud]^2\bar s$ and/or triquarks like $[ud][ud\bar s]$ which
amplify the attractive color-magnetic forces. There are two routes
that emerge naturally. One is that of \cite{Karliner}, based on
the SZ model. The other is the Jaffe-Wilczek (JW) model
\cite{Jaffe:2003}, where it has been proposed that the systematics
of exotic baryons can be explained by diquark correlations.

Neither the SZ model nor the constituent QM have been yet derived
from QCD.  Therefore it is tempting to consider the Effective
Hamiltonian (EH) approach in QCD (see {\it e.g.} \cite{lisbon})
which, from one side, can be derived from QCD and, from another
side, leads to the results for the $\bar q q$ mesons and $3q$
baryons which are equivalent to the quark model ones with some
important modifications. Note that the EH approach contains the
minimal number of input parameters: current (or pole) quark
masses, the string tension $\sigma$, and the strong coupling
constant $\alpha_s$, and does not contain additional fitting
parameters.
It should be useful and attractive to consider expanding this
approach to include diquark degrees of freedom with appropriate
interactions. The preview of this  program based on assumption
that chiral forces are responsible for the formation of $ud$
diquarks while the strings are mainly responsible for binding
constituents in $\Theta^+$  was done in \cite{NTS03}. In this
paper we briefly review application of the EH approach to the JW
model of pentaquarks.

In this model for the $\Theta^+(1540)$ and other $q^{4}\bar q$
baryons the four quarks are bound into two scalar, singlet isospin
diquarks. Diquarks must couple to ${\bf 3}_c$ to join in a color
singlet hadron. In total there are six flavor symmetric diquark
pairs $[ud]^2$, $[ud][ds]_+$, $[ds]^2$, $[ds][su]_+$, $[su]^2$,
and $[su][ud]_+$ which, combining with the remaining antiquark, give
18 pentaquark states in ${\bf 8}_F$ plus ${\bf\overline{10}}_F$.
All these states are degenerate in the SU(3)$_F$ limit. In the QM,
the five quarks are connected by seven strings. In the diquark
approximation
the five-quark system effectively reduces to the three-body one,
studied within the EH approach in \cite{baryons}.
\section{The Effective Hamiltonian Approach}

 The EH for the three constituents has the form
\begin{equation}
\label{EH} H=\sum\limits_{i=1}^3\left(\frac{m_i^{2}}{2\mu_i}+
\frac{\mu_i}{2}\right)+H_0+V,
\end{equation}
where $H_0$ is the kinetic energy operator, $V$ is the sum of the
perturbative one-gluon exchange potential and the string
potential which is proportional to the total length of the
strings, {\it i.e} to the sum of the  distances of (anti)quark or
diquarks from the string junction point. In the Y-shape,
the strings meet at $120^\circ$ in order to insure the
minimum energy. This shape moves continuously to a two--leg
configuration where the legs meet at an angle larger than
$120^\circ$.

The constituent masses $\mu_i$  are expressed in terms of the
current quark masses $m_i$ from the condition of the minimum of
the hadron mass $M_H^{(0)}$ as a function of $\mu_i$
\footnote{~Technically, this is done using the auxiliary field
(AF) approach to get rid of the square root term in the Lagrangian
\cite{brin77}. Applied to the
QCD Lagrangian, this technique yields the EH  for hadrons (mesons,
baryons, pentaquarks) depending on auxiliary fields $\mu_i$. In
practice, these fields are finally treated as $c$-numbers
determined from (\ref{minimum_condition}).}:
\begin{equation} \label{minimum_condition}
\frac{\partial M_H^{(0)}(m_i,\mu_i)}{\partial \mu_i}=0,
\end{equation}
where\begin{equation} \label{M_H^{(0)}}
M_H^{(0)}=\sum\limits_{i=1}^3\left(\frac{m_i^{2}}{2\mu_i}+
\frac{\mu_i}{2}\right)+E_0(\mu_i). \end{equation} In Eq.
(\ref{M_H^{(0)}}) $E_0(\mu_i)$ is an eigenvalue of the operator
$H_0+V$. Quarks acquire constituent masses
$\mu_i\sim\sqrt{\sigma}$ due to the string interaction in
(\ref{EH}). The EH in the form of (\ref{EH}) does not include
chiral symmetry breaking effects. A possible interplay between
these effects should be carefully clarified in the future.

The physical mass $M_H$ of a hadron is
\begin{equation}\label{self_energy}
M_H=M_H^{(0)}+\sum_i C_i.
\end{equation}
The (negative) constants $C_i$ have the meaning of the constituent
self energies and are explicitly expressed in terms of string
tension $\sigma$ \cite{simonov_self_energy} \be\label{c_i}
C_i=-\frac{2\sigma}{\pi\mu_i}\eta_i,\ee where $\eta_q=1$,
$\eta_s=0.88$ is the correction factor due to non-vanishing
current mass of the strange quark. The self-energy corrections are
due to constituent spin interaction with the vacuum background
fields and equal zero for any scalar constituent. The accuracy  of
the EH method for the three-quark systems is $\sim 100$ MeV
\cite{baryons}. One can expect the same accuracy for the
diquark-diquark-(anti)quark system.

Consider a pentaquark consisting of two identical diquarks with
current mass $m_{[ud]}$ and an antiquark with current mass
$m_{\bar q}$ ($q=d,s$). In the hyperspherical formalism the wave
function $\psi( {\rho},{\lambda)}$ expressed in terms of the
Jacobi coordinates $\rho$ and $\lambda$  can be written in a
symbolical shorthand as
\begin{equation}\psi(\rho,\lambda)=\sum\limits_K\psi_K(R)Y_{[K]}(\Omega).
\end{equation} where $R^2={\rho}^2+{\lambda}^2$, and  $Y_{[K]}$ are eigenfunctions (the
hyperspherical harmonics)  of the angular momentum operator $\hat
K(\Omega)$ on the 6-dimensional sphere:
$\hat{K}^2(\Omega)Y_{[K]}=-K(K+4)Y_{[K]}$, with $K$ being the
grand orbital momentum.  The Schr\"odinger equation written in
terms of the variable $x=\sqrt{\mu} R$, where $\mu$ is an
arbitrary scale of mass dimension which drops out in the final
expressions, reads:
\begin{eqnarray} \label{shr}
&&\frac{d^2\chi_K(x)}{dx^2}+\nonumber\\
&&2\left[E_0+\frac{a_K}{x}-b_Kx-\frac{{\cal L}^2(K)}{2x^2}\right]
\chi_K(x)=0,
\end{eqnarray}
with the boundary condition $\chi_K(x) \sim {\cal O} (x^{5/2+K})$
as $x\to 0$ and the asymptotic behavior $\chi_K(x)\sim
{\mathrm{Ai}}((2b_K)^{1/3}x)$ as $x\to \infty$. In Eq. (\ref{shr})
${\cal L}^2(K)=(K+\frac{3}{2})(K+\frac{5}{2})$,
\begin{equation}
a_K=R\sqrt{\mu}\cdot \int V_{C}({\bf r}_1,{\bf r}_2,{\bf
r}_3)\cdot u_K^2\cdot d\Omega,
\end{equation}where
\begin{equation}
V_{C}({\bf r}_1,{\bf r}_2,{\bf r}_3)=
-\frac{2}{3}\alpha_s\cdot\sum\limits_{i<j}\frac{1}{r_{ij}},
\end{equation}and\begin{equation}
b_K=\frac{1}{R\sqrt{\mu}}\cdot\int V_{\rm{string}}({\bf r}_1,{\bf
r}_2,{\bf r}_3)\cdot u_K^2\cdot d\Omega,\label{b_int}
\end{equation}
The explicit expression of
$V_{\rm{string}}({\bf{r}}_1,{\bf{r}}_2,{\bf{r}}_3)$ in terms of
Jacobi variables is given in \cite{plekhanov}.

For \emph{identical} diquarks, like $[ud]^{2}$, the lightest state
must have a wave function antisymmetric under diquark space
exchange. For a state with $L=1,~l_{\rho}=1,~l_{\lambda}=0$ the
wave function in the lowest hyperspherical approximation $K=1$
reads
\begin{equation}
 \psi=R^{-5/2}\chi_1(R)u_1(\Omega),~~
 u_1(\Omega)=\sqrt{\frac{6}{\pi^3}}\sin\theta\cdot
 ({\bf {n}}_{\rho})_z.
\end{equation} Here, one unit of orbital momentum between the
diquarks is introduced with respect to
the $\bf\rho$ variable whereas the ${\bf\lambda}$ variable is in
an S-state.

In what follows, we use $\sigma=0.15$~GeV$^2$, and explicitly
include the Coulomb--like interactions between diquarks and
antiquark with $\alpha_s=0.39$.

The mass of the $\Theta^+$ obviously depends on $m_{[ud]}$ and
$m_s$. The current masses of the light quarks are relatively
well-known: $m_{u,d}\approx 0$, $m_s\approx 170$ MeV. The only
other parameter for strong interactions is the current mass of the
diquark $m_{[ud]}$. In principle, this mass could be computed
dynamically. Instead, one can tune $m_{[ud]}$ (as well as
$m_{[us]}$ and $m_{[ss]}$) to obtain the baryon masses in the
quark-diquark approximation. We shall comment on this point later
on.

\section{The results}
For pedagogy, let us first assume  $m_{[ud]}=0$. This
assumption leads to the lowest $uudd\bar d$ and $uudd\bar s$
pentaquarks. If the current diquark masses vanish, then the
$[ud]^2\bar d$ pentaquark is dynamically exactly analogous to the
$J^P=\frac{1}{2}^-$ anti-nucleon resonance and $[ud]^2\bar s$
pentaquark is an analogue of  the $J^P=\frac{1}{2}^-$
anti-$\Lambda$ hyperon, with one important exception. The masses
of P-wave baryons calculated using the EH method acquire the
(negative) contribution $3C_q$ for $N(\frac{1}{2}^-)$ and
$2C_q+C_s$ for $\Lambda (\frac{1}{2}^-)$ due to the constituent
spin interaction with the vacuum chromomagnetic field.

However, for the {\it scalar diquarks} the above discussion shows
that the self-energies $C_{[ud]}$ equal zero . This means that
introducing any scalar constituent increases the pentaquark energy
(relative to the $N$ and $\Lambda$ P-wave resonances by
$2|C_q|\sim 200-300$ MeV. Therefore prior to any calculation we
can put the lower bound for the pentaquark in the JW
approximation: $M(\Theta)\ge 2~ {\rm GeV}$.

The numerical calculation for $m_{[ud]}=0$ yields the mass of the
$[ud]^2\bar s$ pentaquark $\sim$ 2100 MeV (see Table \ref{EH for
pentas}). Similar calculations yield the mass of the $[ud]^2\bar
c$ pentaquark $\sim$ 3250 MeV (for $m_c=1.4$ GeV,~$\eta_c=0.234$)
and the mass of the $[ud]^2\bar b$ pentaquark $\sim$ 6509 MeV (for
$m_b=4.8$ GeV, $\eta_b=0.052$) \cite{veselov}. For illustration of
the accuracy of the AF formalism, in Table \ref{EH for pentas} are
also shown the masses of $[ud]^2\bar s$ and $[ud]^2\bar d$
pentaquarks calculated using the spinless Salpeter equation (SSE)
corresponding to the relativistic Hamiltonian
:\begin{eqnarray}\label{SSE} H_S &=&
\sum_{i=1}^3 (\vec p_i\,^2+ m_i^2)^{\frac{1}{2}} + V, \nonumber \\
M &=& M_0 -\frac{2\sigma}{\pi} \sum_{i=1}^3 \frac{\eta_i}{\langle
\sqrt{\vec p_i\,^2+ m_i^2 \rangle}},
\end{eqnarray}
where $V$ is the same as in Eq. (\ref{EH})~\footnote{The numerical
algorithm to solve the three-body SSE is based on an expansion of
the wave function in terms of harmonic oscillator functions with
different sizes \cite{nunb77}. In fact to apply this techniques to
the three-body SSE we need to use an approximation of the
three-body potential $V_{{\rm string}}$ by a sum of the two- and
one-body potentials, see \cite{NSSB}. This approximation, however,
introduces a negligible correction to energy eigenvalues.} and
$M_0$ is the eigenvalue of $H_S$. $\mu_{[ud]}$ and $\mu_q$ in
Table 1 denote the constituent masses calculated in the AF
formalism using Eq. (\ref{c_i}) or the mean values $\langle({\vec
p}_{[ud]}\,^2+m_{[ud]}^2)^{\frac{1}{2}}\rangle$, $\langle({\vec
p}_{\bar s}\,^2+m_{\bar s}^2)^{\frac{1}{2}})\rangle$ found from
the solution of SSE.

\begin{table} \caption{The pentaquark masses in the
diquark-diquark-antiquark approximations.  Shown are the masses of
$[ud]^2\bar{q}$ states ($q=d,s$) for $J^P=\frac{1}{2}^+$
pentaquarks. }
\begin{center}
{\renewcommand{\arraystretch}{0}%
\begin{tabular}{|c|c|c|c|c|}
\hline
\strut \rule[-10pt]{0pt}{25pt}& &\(\mu_{[ud]}\) & \(\mu_q\) & \(M\) \\
\hline
\rule{0pt}{2pt} & & & & \\
\hline \strut \rule[-10pt]{0pt}{28pt}
\([ud]^2\bar{s}~\frac{1}{2}^+\) & AF & 0.467 & 0.470 & 2.115 \\
\cline{2-5} & SSE & 0.463 & 0.468 & 2.070 \rule[-10pt]{0pt}{28pt}\\
\hline
\rule{0pt}{2pt} & & & & \\
\hline \strut \rule[-10pt]{0pt}{28pt}
\([ud]^2\bar{d}~\frac{1}{2}^+\) & \strut AF & 0.491 & 0.415& 2.065 \\
\cline{2-5} & SSE & 0.469 & 0.379 & 1.934 \rule[-10pt]{0pt}{28pt}\\
\hline
\end{tabular}
}
\end{center}
\label{EH for pentas}
\end{table}

\noindent Another way to estimate the current diquark masses  is
to tune $m_{[ud]}$, $m_{[us]}$ and $m_{[ss]}$ from a fit to the
nucleon and hyperon masses (in the quark-diquark approximation).
We have performed such calculations using the SSE. We briefly
investigated the sensitivity of the pentaquark mass predictions to
the choice of the diquark masses
and found that the choice $m_{[ud]}=455$ MeV, $m_{[us]}=565$ MeV,
and $m_{\bar s}=170$~MeV is quite acceptable. For this choice we
obtain $M_{[ud]u}= 1026~{\rm MeV},$ ~$M_{[ud]s}= 1116~{\rm MeV},$
$M_{[us]d}=1118~ {\rm MeV},$~$M_{[us]s}= 1206~ {\rm MeV}$, yielding
for the pentaquark masses
\begin{eqnarray}\label{choice_1}&&M([ud]^2\bar d)= 2377~{\rm
MeV},\nonumber\\&&M([ud]^2\bar s)= 2435~{\rm
MeV},\nonumber\\&&M([us]^2\bar d)= 2533~{\rm MeV},\end{eqnarray}
These values are typically $\sim 400$ MeV higher the lowest bounds
reported in Table \ref{EH for pentas}. For illustration, we
mention that another choice $m_{[ud]}=285~ {\rm MeV}$,
$m_{[us]}=515~ {\rm MeV}$, $m_{\bar s}=260~{\rm MeV}$ yields
$M_{[ud]u}= 941~{\rm MeV}$, $~M_{[ud]s}= M_{[us]d}=1117 ~{\rm
MeV},$~$M_{[us]s}= 1282~ {\rm MeV},$ and the pentaquark masses
\begin{eqnarray}\label{choice_2}&&M([ud]^2\bar d)= 2288~{\rm
MeV},\nonumber\\&&M([ud]^2\bar s)= 2408~{\rm MeV},\nonumber\\&&
M([us]^2\bar d)= 2573~{\rm MeV},\end{eqnarray} that agree within
$\sim 100$ MeV with (\ref{choice_1}).

Since two diquarks are spin and isospin singlets, there is no
contribution from the spin part of one-gluon exchange potentials
and spin-isospin part of Goldstone boson exchange potentials
between the two diquarks and each diquark and the $\bar s$ quark.
Only the $\sigma$ meson exchange potential  produces the decrease
of the pentaquark binding energy; this effect is estimated by no
more than $\sim$ 100 MeV. The instanton induced spin interaction
also produces a marginal attraction \cite{ss}. We therefore
conclude that the string dynamics alone in its simplified form
predicts too high masses for pentaquarks.

\section{Conclusions}

The anomalous lightness of the $\Theta^+$ seems to unambiguously
indicate a large role of chiral symmetry breaking effects
in pentaquarks.
The QM and $\chi$SM are, to a large extent, complementary. Each of
these reproduces certain aspects of hadronic physics and
incorporates some features of QCD that are missing in the other.
An approach of $\chi$SM totally neglects the confinement effects
and concentrates on the pure chiral properties of baryons. The
string dynamics alone, in its simplified form, seems to predict
too high masses for pentaquarks. Therefore the existence of
$\Theta^+$, if confirmed, provides  a unique possibility to
clarify the interplay between the gluon and chiral degrees of
freedom in light baryons.

\section*{Acknowledgement}
\noindent This work was supported by RFBR grants 03-02-17345,
04-02-17263, 04-02-26642  and the grant for leading scientific
schools  1774.2003.2.

\end{document}